\documentclass[12pt]{article}
\pdfoutput=1
\usepackage{graphicx}
\usepackage{amssymb} \usepackage{amsmath} \usepackage{esint}
\usepackage{bm}

\usepackage{bm}

\newcommand{\rf}[1]{(\ref{#1})} \newcommand{\beq}{\begin{equation}}
 \newcommand{\eeq}{\end{equation}}
\newcommand{\bea}{\begin{eqnarray}} \newcommand{\eea}{\end{eqnarray}}

 \newcommand{\Del}{\Delta}
\newcommand{\sg}{\sigma}

\newcommand{\prt}{\partial}

\begin{document}

\begin{center} \vspace{24pt} {\large \bf The spectral dimension in 
2D CDT gravity coupled to scalar fields}
\vspace{30pt}

{\sl J. Ambj\o rn}$\,^{a,b}$, {\sl A. G\"{o}rlich}$\,^{a,c}$ {\sl J.
Jurkiewicz}$\,^c$ and {\sl H. Zhang}$\,^c$

\vspace{48pt}

$^a$~The Niels Bohr Institute, Copenhagen University\\ Blegdamsvej 17,
DK-2100 Copenhagen \O , Denmark.\\ email: ambjorn@nbi.dk,
goerlich@nbi.dk, zhang@if.th.uj.edu.pl\\

\vspace{10pt}

$^b$~Institute for Mathematics, Astrophysics and Particle Physics
(IMAPP)\\ Radbaud University Nijmegen, Heyendaalseweg 135, 6525 AJ, \\
Nijmegen, The Netherlands

\vspace{10pt}

$^c$~Institute of Physics, Jagiellonian University,\\ Reymonta 4, PL
30-059 Krakow, Poland.\\ email: jerzy.jurkiewicz@uj.edu.pl\\

\vspace{96pt} \end{center}

%
%
\begin{center} {\bf Abstract} \end{center}

\noindent Causal Dynamical Triangulations (CDT) provide a
non-perturbative formulation of Quantum Gravity assuming the existence
of a global time foliation. In our earlier study we analyzed the effect
of including $d$ copies of a massless scalar field in the two-dimensional
CDT model with imaginary time. For $d > 1$ we observed the formation of a
``blob", somewhat similar to that observed in four-dimensional CDT
without matter. In the two-dimensional case the ``blob" has a Hausdorff
dimension $D_H=3$. In this paper we study the spectral dimension $D_S$ of the
two-dimensional CDT-universe, both for $d = 0$ (pure gravity) and $d = 4$.
We show that in both cases the spectral dimension is consistent with
$D_S = 2$.

\newpage

\section{Introduction}\label{Introduction}

Quantizing gravity remains one of  big challenges for Theoretical Physics. 
Causal Dynamical Triangulations (CDT) is  one of a number 
of attempts to define non-perturbatively a theory of 
quantum gravity. We refer to \cite{physrep} for a recent review.  
In CDT spacetime is discretized (triangulated) in such a way that 
a global time foliation exists. This time foliation permits one to 
perform a Wick rotation to imaginary time for each configuration,
and it makes it possible to study the CDT spacetimes
using standard methods of statistical mechanics, i.e.\ Monte Carlo
simulations.

The existence of a global time foliation has important consequences.
The regularized theory of quantum gravity defined this way differs 
from the regularized theory without this constraint. If the constraint
of a time foliation is not imposed and we use triangulations with 
Euclidean signature, the regularization is denoted Euclidean dynamical 
triangulation (EDT). In four dimensions EDT seems not to result in an
interesting continuum limit when the UV cut off (the link length in the 
triangulations) is taken to zero. That was the reason for inventing the 
formalism of CDT, which might indeed have an interesting continuum
limit in the case of the three-- and four--dimensional theories, 
\cite{3d,4d-1,4d-2,4d-3}.
The  regularized two-dimensional theories can be solved analytically 
both for CDT and for EDT \cite{ambjorn,david,kkm} and the continuum limit
corresponds to 2d Ho\v{r}ava-Lifshitz quantum gravity \cite{horava,hlqg} 
and $c_L=26$ Liouville quantum gravity \cite{kpz,david2,dk}, respectively,
($c_L$ denoting the central charge of the Liouville theory).
If one couples matter to the regularized 2d quantum gravity theories  
one can explicitly solve modified EDT models which in 
the continuum limit correspond to rational (p,q)-minimal
conformal field theories with central charge $c_M \leq 1$ coupled to 2d 
Liouville quantum gravity with corresponding Liouville central charge 
$c_L=26-c_M$. When $c_M >1$ the combined regularized theory seeming
degenerate to a theory of branched polymers. This is the so-called 
$c_M =1$ barrier. It has not been possible to solve 
analytically CDT coupled to matter fields except in a 
few cases where the relation to a central charge of the 
matter fields is not so clear \cite{unclear}. However, one 
can study CDT coupled to matter fields with well-defined central 
charges numerically. Ising models and Potts models  coupled to CDT, as 
well as Gaussian scalar fields coupled to CDT have been 
studied using Monte Carlo simulations \cite{cdtmatter1,cdtmatter2}.

In these Monte Carlo simulations the following was observed:
for $c_M \leq 1$ the interaction between matter 
and geometry is weak in the sense 
that the critical exponents of the matter fields are the same as
the exponents of matter fields in flat spacetime (this is in contrast to 
the critical exponents of matter in EDT). At the same time the Hausdorff
dimension $d_H$ of the CDT spacetime is 2 if the  
matter fields have central charges $0\leq c_M \leq 1$ 
(again in contrast to the Hausdorff dimension of EDT spacetime). 
However, like in EDT there 
seems to be a $c_M =1$ barrier. The EDT $c_M=1$ barrier can be viewed as
the proliferation of baby universes such that the EDT geometries
degenerate into so-called branched polymers. The imposed time foliation
in CDT prevents such a generation of baby universes. Instead the 
$c_M=1$ barrier manifests itself by the generation of a ``blob''.
By a ``blob'' we mean the following: in the computer simulations 
we choose a certain volume of spacetime, i.e.\ the number of triangles $N$.
Also we choose a certain number of time-foliations $L$. For 
each lattice time $t \in [0,L]$ there is associated a space-volume
$n(t)$. However, for a fixed $N$ and if  $L$ is sufficiently large, 
the space volume $n(t)$ is only significantly different 
from zero in some sub-interval 
of $L$. This defines the extent of the blob, and this  distribution of $n(t)$
is in sharp contrast with the situation for $c_M \leq 1$ where $n(t)$ is 
uniformly distributed over the whole range $L$ of times. Moreover
the time extension of the blob seems to  scale universally as $N^{1/3}$
no matter which $c_M > 1$ is used and which kind of critical matter 
(Gaussian fields, multiple Ising spins etc.) one uses. This scaling 
behavior indicated that the Hausdorff dimension $D_H$ of the blob is 3.
The formation of a ``blob'' has been observed also in higher dimensional
CDT where it has been identified with a ``de Sitter'' like phase of 
the quantum universe of the CDT model \cite{semiclassical,4d-2}.    

To further understand the geometries generated by the CDT models,
we measure in this article the spectral dimension $D_S$ of two-dimension CDT, 
both for $c_M=0$ and for $c_M >1$  where we have the dynamical generation of 
a ``blob''. We use the diffusion equation on the CDT configurations
to determine the spectral dimension and we find $D_S=2$ in all cases.
The topology of the spacetime manifold we use is by construction 
that of a torus. However, as described above, when a blob is formed
we have effectively a change in topology from a torus to a sphere  
since the thin ``stalk'' which constitutes spacetime outside the blob
carries almost no spacetime volume. We show that the
standard method used to  measure $D_S$ can be used to discriminate between 
the two if we use suitable long diffusion time.

In Sec.\  \ref{model} we briefly recall the definition of 
the 2d CDT model and the geometric
scaling used in the case of pure gravity and gravity interacting with
matter fields. In the Sec.\  \ref{spectral} 
we shortly discuss the behavior of the
spectral dimension defined using the return probability of the diffusion
process on a regular two-dimensional surface with a topology of a torus
and of a sphere. We use these results to analyze results of numerical
simulations in the cases $c_M=0$ and $c_M=4$ (four Gaussian
fields coupled to geometry). The last section contains
conclusions and discussion.

\section{Definition of the 2d CDT  model}\label{model}

The CDT model in 1+1 dimensions  represents the path 
integral as a sum over 
piecewise linear spacetimes built of triangles \cite{ambjorn}. The vertices of
triangles are parametrized by an integer time variable, 
with two vertices at  time
$t$ and one vertex at $t\pm 1$. Spatial slices at integer time have the topology
of the circle $S_1$. An alternative formulation can be based on the dual
lattice with points in the centers of the triangles, connected by links, dual
to the links of the original triangles. The dual vertices can be viewed
as placed at  half-integer times. In the dual formulation each vertex is
connected to three other vertices, two in the same time slice and one
above or below at half-integer time. Links joining vertices with the
same time index form a circle. The dual
formulation is completely equivalent to the one formulated in terms 
of triangles and will be used in this paper. 
In \cite{ambjorn} the starting point was Lorentzian spacetime,
where two of the links of each triangle had time-like signature.
However, a Wick rotation could be performed for each triangulation
such that all triangles became ordinary Euclidean equilateral 
triangles where all links have  lengths  $a$. We work here with 
this Wick rotated CDT version. In
practice we take $a = 1$ and postpone the discussion of the continuum
limit $a \to 0$ until later. In this formulation we map the quantum
system into a statistical system, described by the partition function
\begin{equation} 
Z = \sum_T \int' \prod_{i,\mu} d\phi_i^{\mu}\;
\frac{1}{C_T}\;\exp\left(-\lambda N_T -
S_{M}(\phi_i^{\mu})\right) 
\label{partition} 
\end{equation} 
where $\lambda$ is a cosmological constant, $N_T$ is the number of vertices in
a (dual) triangulation $T$ and  $C_T$ is a symmetry factor of the
triangulation. 
$S_M(\phi)$ denotes the matter action coupled to geometry. We
choose here $d$ Gaussian scalar fields  $\phi_i^{\mu},~\mu=1,\dots, d$,
which thus represent a conformal field theory with 
central charge $c_M = d$. The discretized Gaussian action,
suitable for the (dual) triangulation is 
\begin{equation} 
S_{M} = \frac{1}{2} \sum_{i \leftrightarrow j, \mu}(\phi_i^{\mu}-\phi_j^{\mu})^2 
\label{action} 
\end{equation} 
The summation is over all neighboring pairs of vertices. 
The action is invariant under translation $\phi_i \to \phi_i+\phi_0$,
and we fix this zero-mode in the  action \rf{action}
by fixing the "center of mass" of $\phi$. This is indicated by the 
prime-label in $\int$. $d=0$ corresponds to pure CDT gravity
where $c_M=0$. The requirement that the universe (the dual triangulation)
does not consist of disconnected components implies
that at each time slice of the dual lattice
we have at least one link pointing back in time and one link
pointing forward. For convenience of the computer simulations
and in order to minimize boundary effects we 
choose periodic boundary conditions in the time directions
and we fix the number of integer time steps to be $L$.
The partition function (\ref{partition}) can be expressed as a sum 
\begin{equation} 
Z= \sum_N e^{-\lambda N}\; Z_N, 
\end{equation} 
where in $Z_N$ we sum over  triangulations with $N$ vertices.
A simple quantity we can measure in the computer simulations is 
the spatial volume distribution as a function of time: $n(t)$.
For two-dimensional spacetime $n(t)$ will be a length, and in our 
(dual) triangulation it will be the number of vertices at time $t$.
If a given triangulation has $N$ vertices we obviously have  
$\sum_t n(t) = N$. As mention above it is precisely 
$n(t)$ which signals the $c_M=1$ transition
($n(t)$ changes from an uniform distribution to a ``blob''-distribution).

\begin{figure}[h] \begin{center}
\includegraphics[width=0.45\textwidth]{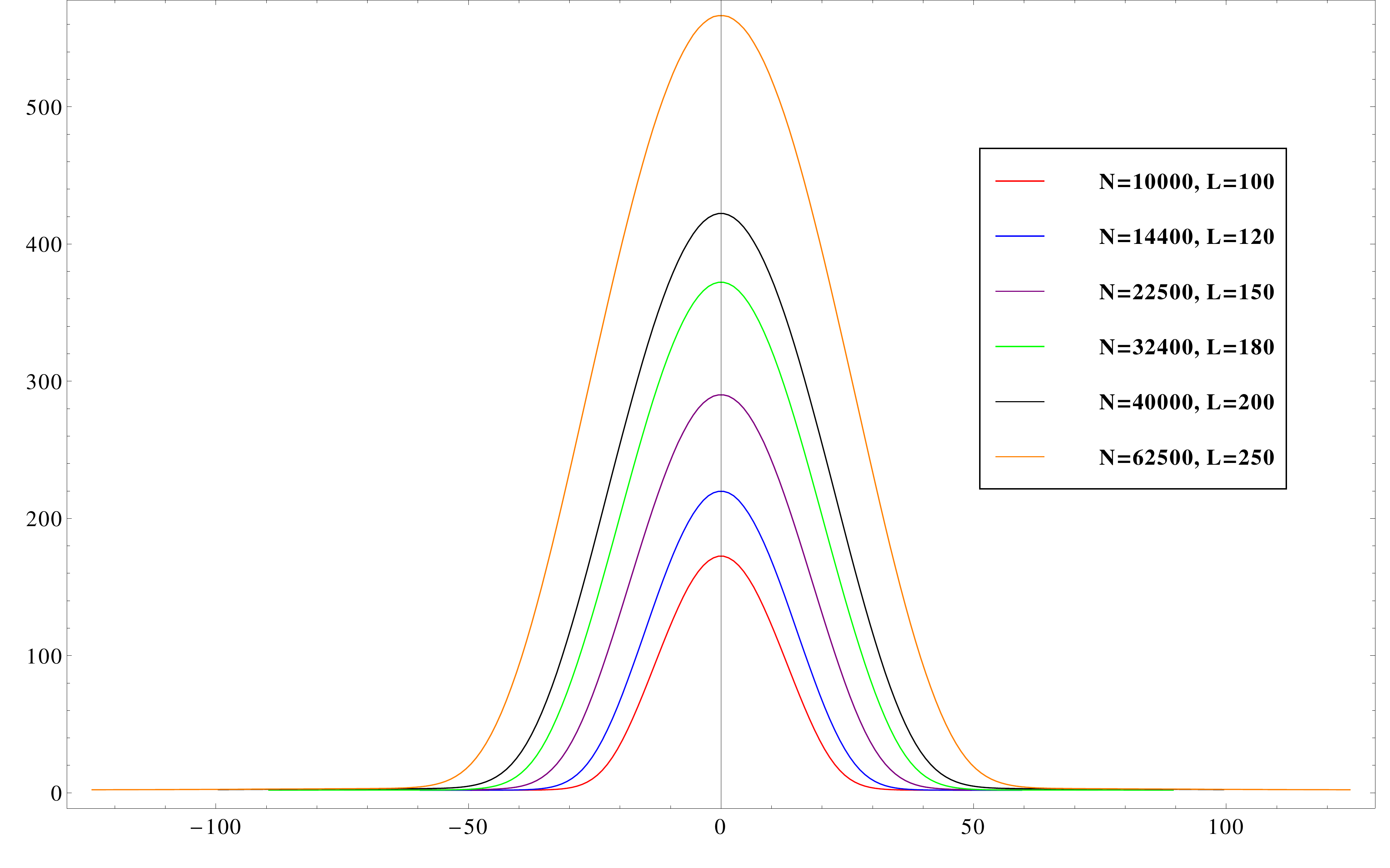}
\includegraphics[width=0.45\textwidth]{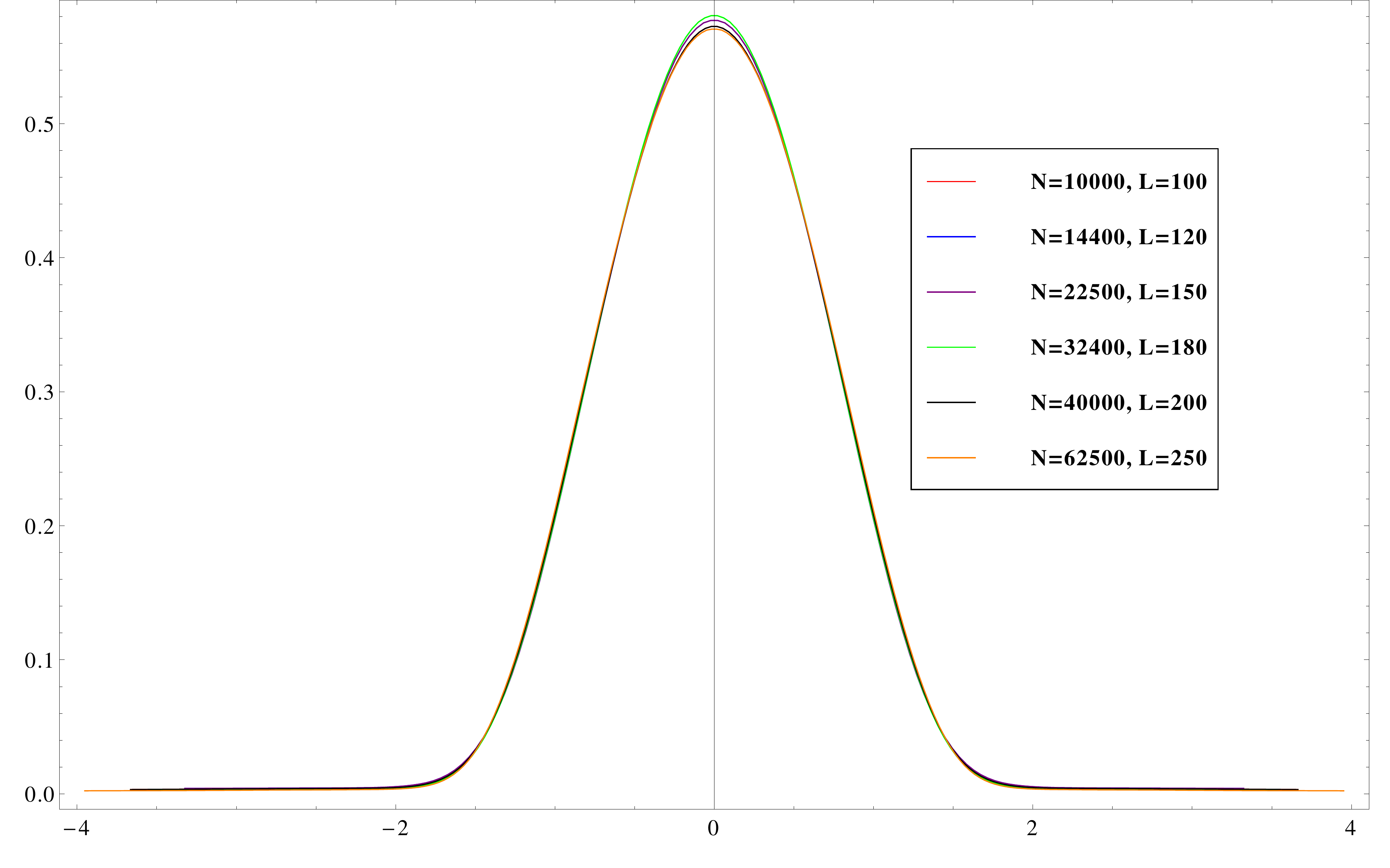}
\end{center} 
\caption{Average volume profile for $d = 4$ (left).
Universal scaled profile, assuming Hausdorff dimension $D_H = 3$
(right).} 
\label{scaling-nt-massless} 
\end{figure}

For the model containing scalar matter fields an analytic solution is
not known but the model can readily be studied using numerical
methods. When performing Monte Carlo simulations it is 
convenient to (approximately) fix the spacetime volume $N$ (the number
of vertices). In order to do that we  use a modified
partition function
\begin{equation}\label{j1} 
Z_N= \sum_{N_T} \exp\left(-\lambda N_T\right) 
Z_{N_T}\exp\left(-\frac{\epsilon}{2}(N_T-N)^2\right)
\end{equation} 
where $\epsilon$ controls the size of spacetime volume fluctuations.
This technique was used successfully in \cite{cdtmatter2} where models
with $d \geq 1$ were studied. What we observed by measuring 
$n(t)$ using the partition function \rf{j1} was the 
formation of a ``blob'' for $d >1$, very similar to what
happens in higher dimensional CDT, as mentioned above. 
Fig.\  \ref{scaling-nt-massless} presents such blob-profiles
$n_N(t)$ for the case of four Gaussian fields ($d=4$). The left
figure shows the average\footnote{The spatial volume profile
for a single triangulation is not a nice curve. However, taking
the average of many such measurements, where we align the ``center of
mass'' of the blobs before taking the average, produces the
nice regular curves shown in the left figure. See \cite{cdtmatter2} 
for details.}  
spatial volume profile $n_N(t)$ for different values of the 
spacetime volume $N$. The right figure shows that all these curves
can be scaled to form a universal curve 
$n_U(t_U)= N^{1-1/D_H} n_N(t/N^{1/D_H})$
with $D_H=3$ being the the Hausdorff dimension of the blob.

\begin{figure}[h] 
\begin{center}
\includegraphics[width=0.45\textwidth]{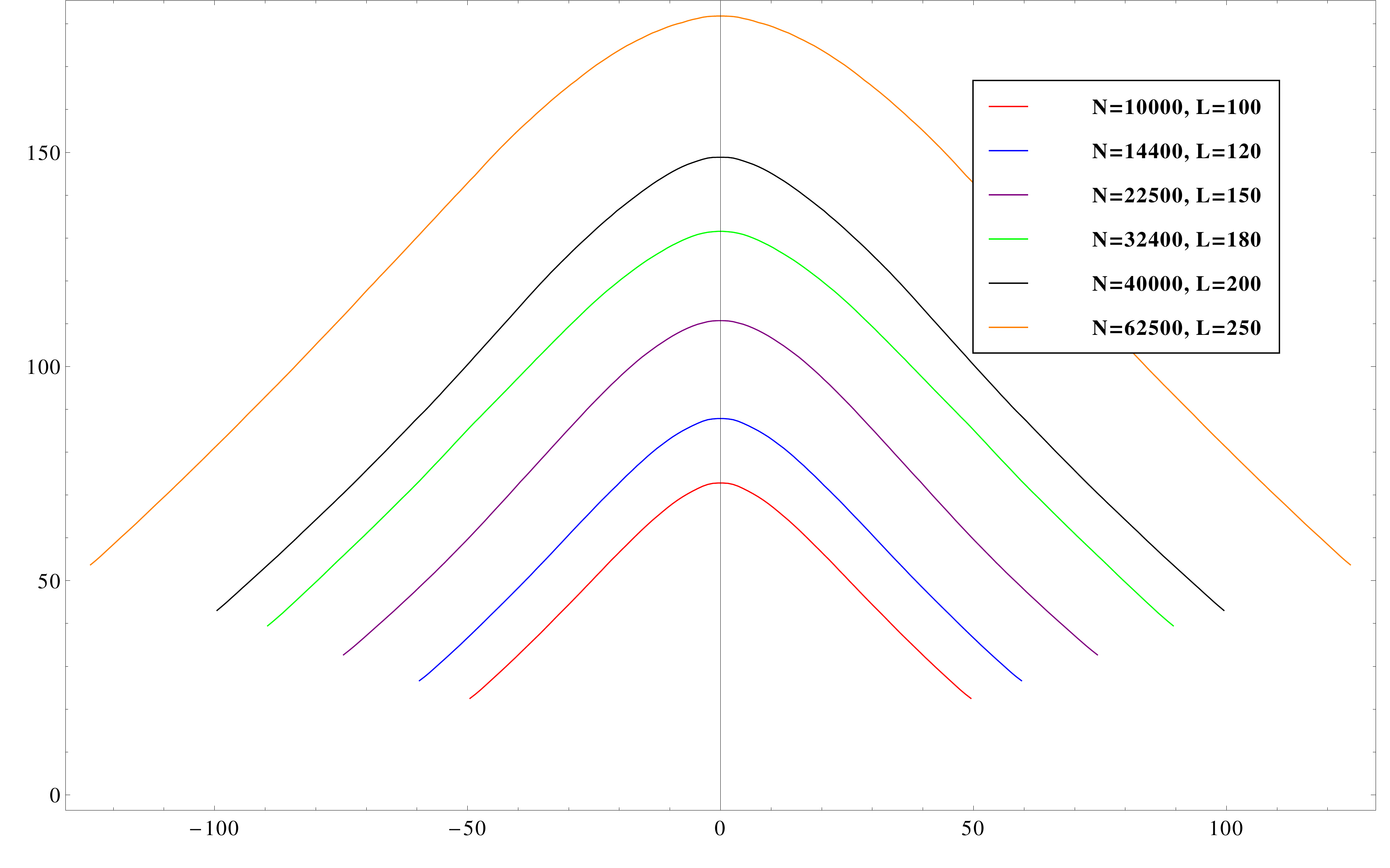}
\includegraphics[width=0.45\textwidth]{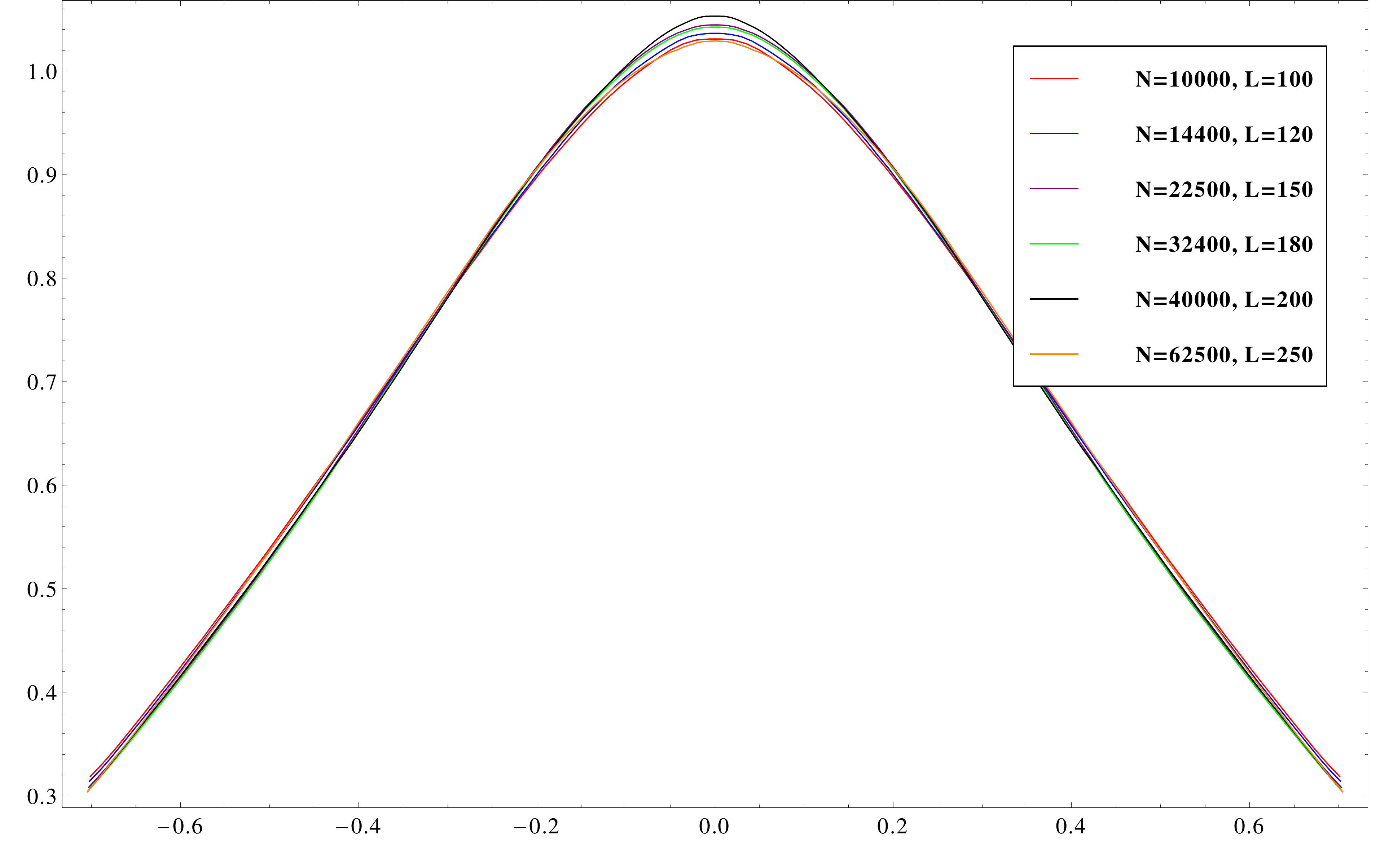} 
\end{center}
\caption{Average volume profile for pure gravity (left) and it's scaled
version (right).} 
\label{scaling-pure} 
\end{figure}

The behavior shows that matter plays a very
important role, inducing a quite nontrivial scaling of the geometry
even for this simple model.
When $N \to \infty$  the ``stalk", the spacetime outside the blob,
effectively disappears and  it is only present because our 
computer algorithm does not allow the spatial volume at a given time
to be zero. The creation of a ``blob'' with non-trivial scaling 
properties and a corresponding stalk should be compared to the 
situation where $c_M \leq 1$. In Fig. \ref{scaling-pure}
we present similar measurements for the case $d=0$.
The $n(t)$'s  on the plot have an
artificial maximum around  time $t=0$, because our procedure
of superimposing different measured configurations defines a
``center of mass'' and allocates time $t=0$ to this center 
of mass before making the superposition. In this way we 
explicitly break the time-translation invariance. 
One observes that no stalk (and thus no blob) is formed. As before
the spatial volume profiles $n_N(t)$ can be scaled to a universal 
profile $n_U(t_U)$, but this time with the canonical dimension
$D_H=2$ which one would naively expect.

\section{The spectral dimension}\label{spectral}

On a fixed, smooth geometry  the
diffusion process is governed by the diffusion equation 
\begin{equation}\label{j2}
\frac{\prt}{\prt \sg} \, K_g(\xi,\xi_0;\sg) = \Del_g K_g(\xi,\xi_0;\sg), 
\end{equation} 
where $\sg$ is the fictitious diffusion
time and $\Del_g$ the Laplace operator corresponding to the metric
$g_{ab}(\xi)$.  $K_g(\xi,\xi_0;\sg)$ is the probability density of
diffusion from $\xi_0$ to $\xi$ in a diffusion time $\sg$. 
The return probability is defined as $K(\xi_0,\sg)=K_g(\xi=\xi_0;\sg)$
and one has the following behavior for short diffusion time $\sg$:
\beq\label{dif1}
K(\xi_0,\sg) \sim \frac{1}{\sg^{D_S/2}} (1+ O(\sg)),
~~~D_S = -2 \lim_{\sg \to 0} \log K(\xi_0,\sg).
\eeq
where $D_S$ is denoted the spectral dimension and for a
smooth manifold is equal to the dimension $D$ of the manifold.
The correction $O(\sg)$ has an asymptotic power expansion 
in $\sg$ (the heat kernel expansion) where the coefficients
depend on the local curvature and derivatives of the curvature at
$\xi_0$.  

\begin{figure}[h] 
\begin{center}
\includegraphics[width=0.45\textwidth]{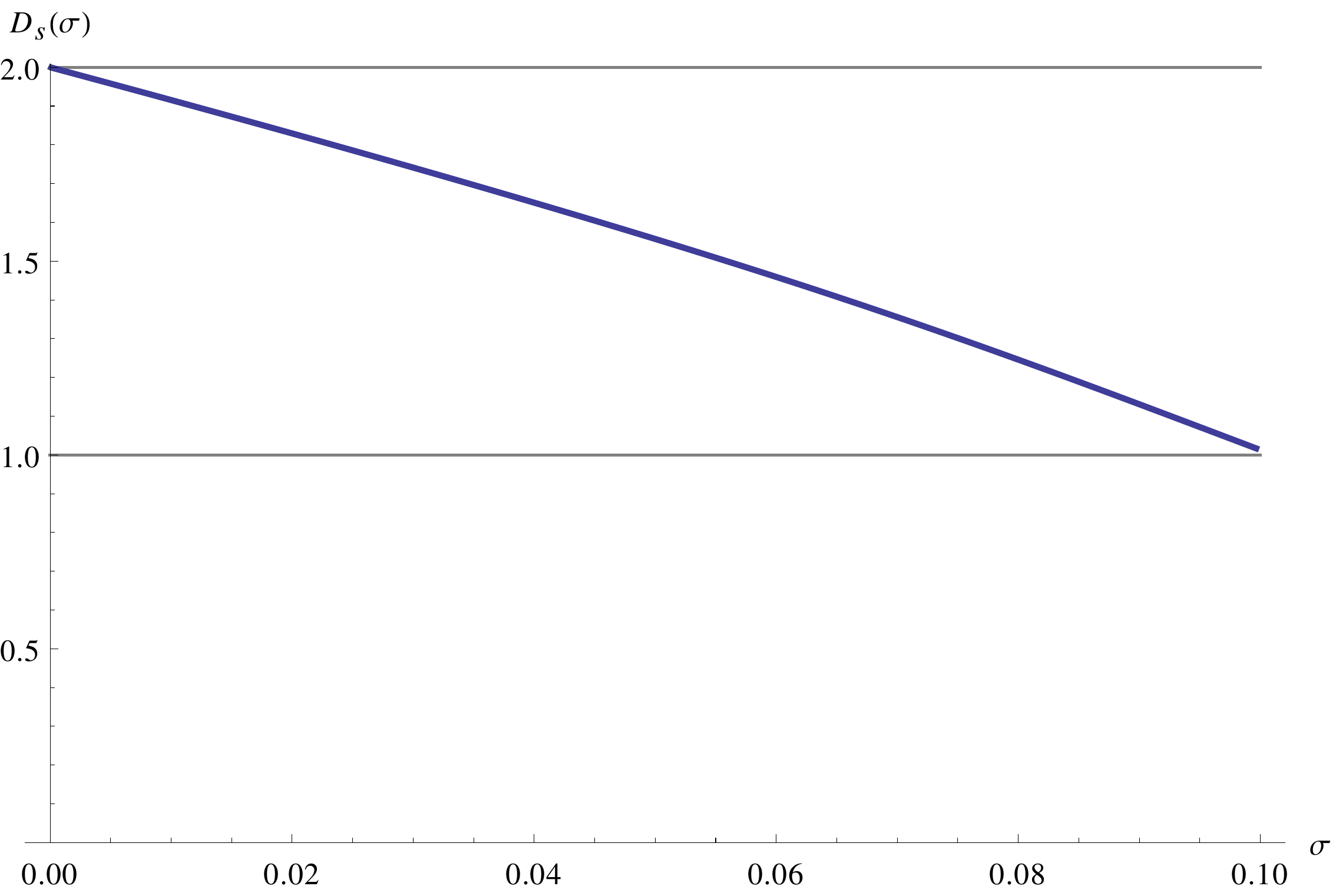}
\includegraphics[width=0.45\textwidth]{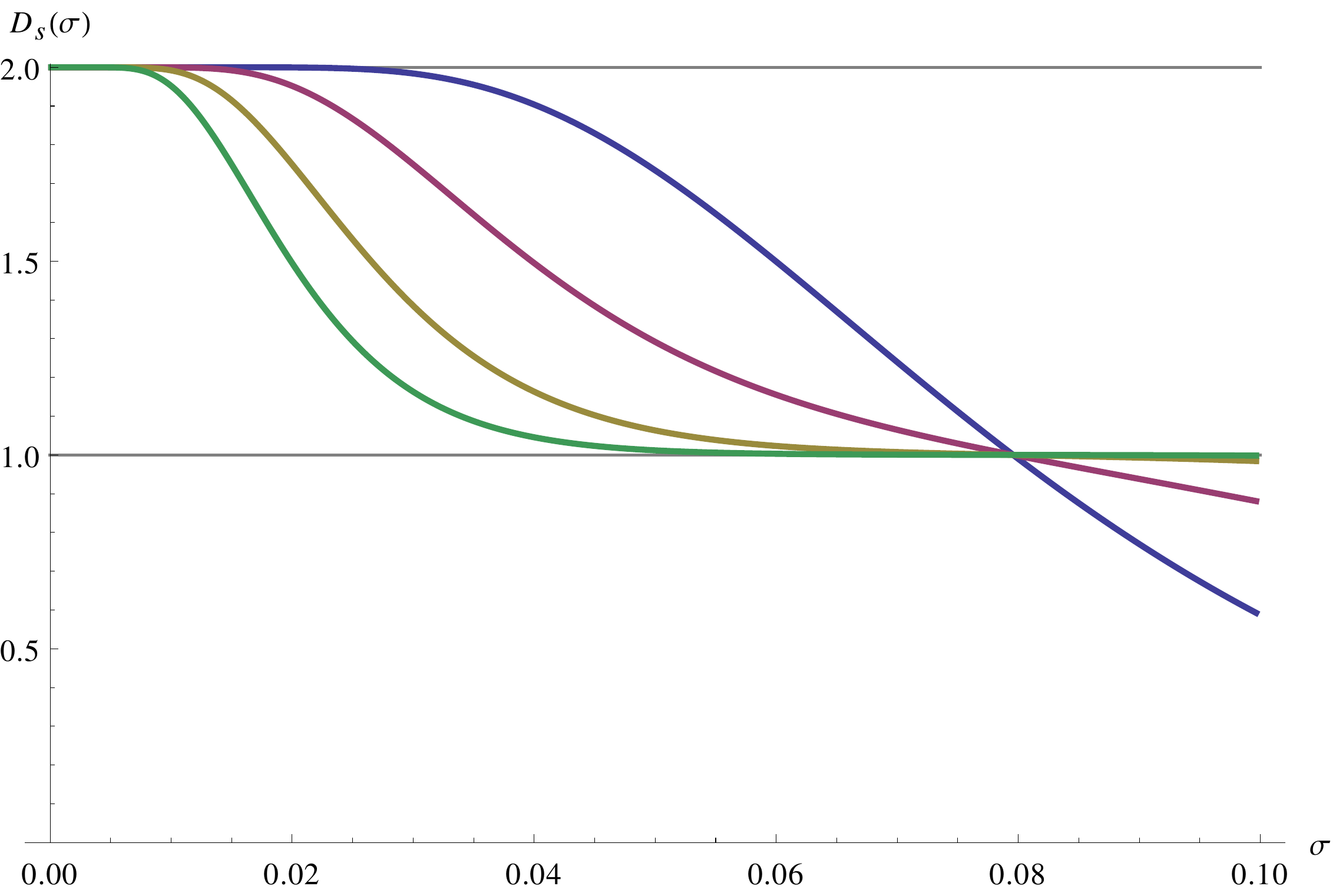}
\end{center} 
\caption{Behavior of $D_S(\sigma)$ for a sphere (left) and
for a torus (right). 
In all cases the area of a manifold is $1$.
For a torus we show  curves for $\rho=1, 2, 3, 4$ and $R^2 = 1$.} 
\label{sp-theory-first}
\end{figure}

\begin{figure}[h] 
\begin{center}
\includegraphics[width=0.47\textwidth]{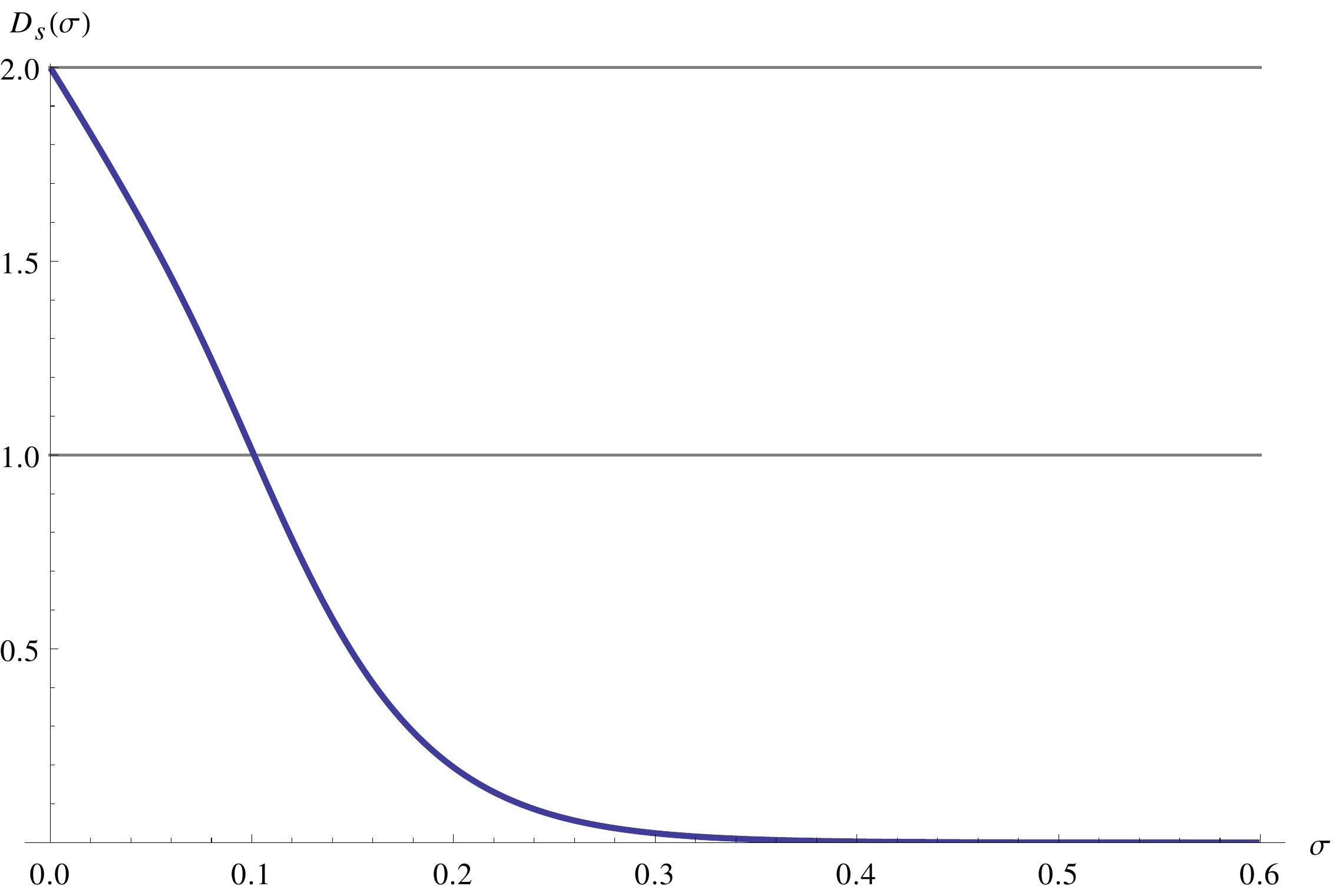}
\includegraphics[width=0.47\textwidth]{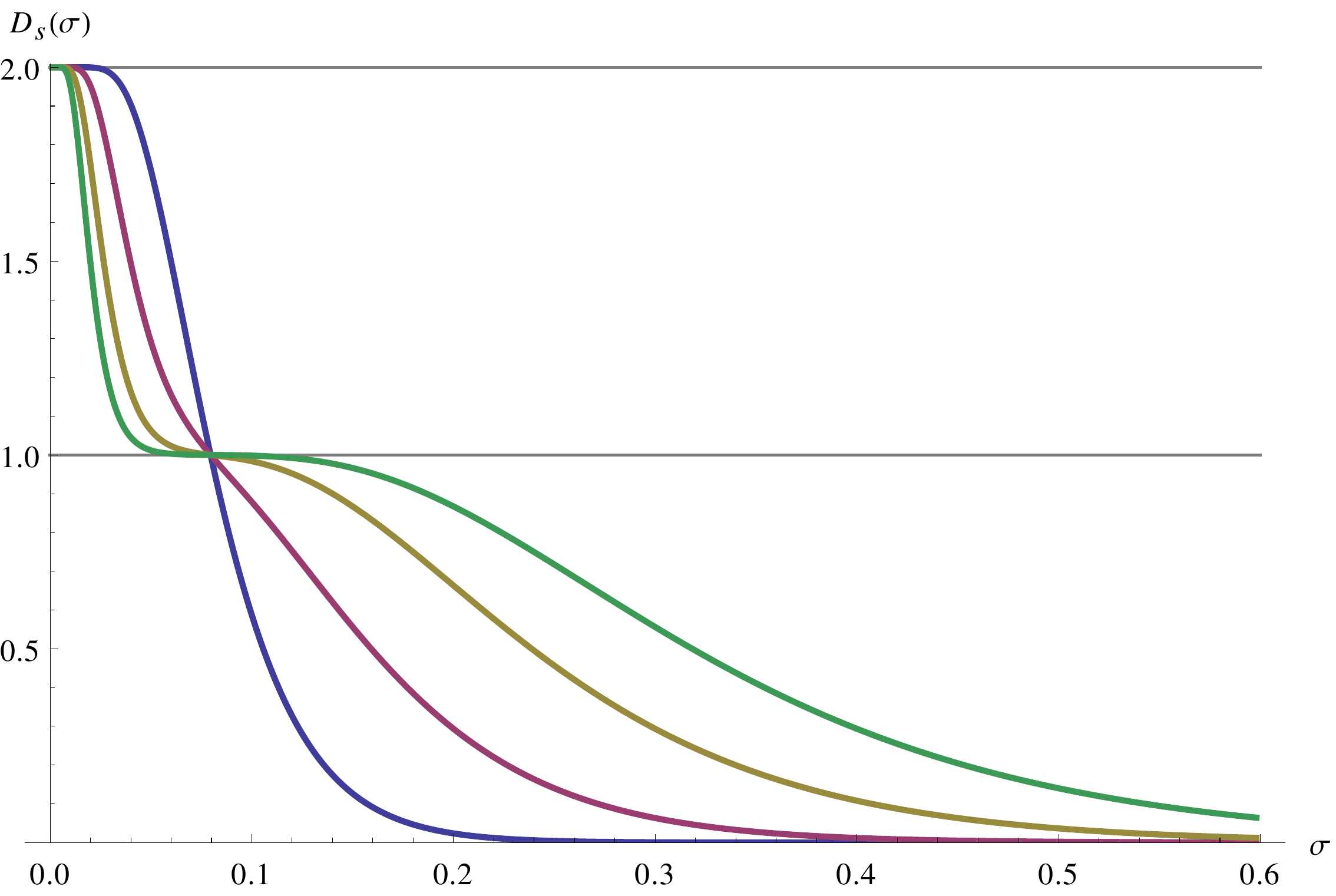}
\end{center} 
\caption{$D_S(\sg)$ for a sphere (left) and a torus (right)
for large $\sg$. In the right plot we
compare the behavior for a torus with $\rho=1,~2,~3$ and 4.}
\label{sp-torus-fix} 
\end{figure}

Let us consider the return probability in two cases: 
a two-dimensional sphere
with the radius $R$ and  a two-dimensional torus with the radii $R_1$
and $R_2$ with $R_1R_2 = R^2$ and $R_1/R_2 = \rho$. In both cases the 
$K(\xi_0,\sg)$ does not depend on the initial position $\xi_0$ and 
we write $K_g(\xi=\xi_0;\sg)=K(\sg)$.  For the sphere
we have
\begin{equation} 
K(\sigma) \propto\sum_{n=0}^{\infty} (2n+1) e^{-n(n+1)\sigma / R^2 }, 
\label{th-spsphere}
\end{equation} 
while the expression for the torus is 
\begin{equation} 
K(\sigma)\propto (1+2\sum_{p=1}^{\infty} e^{-\frac{4 p^2 \pi^2 \sigma
}{R^2\rho}})(1+2\sum_{q=1}^{\infty}e^{ -\frac{4 q^2 \pi^2 \sigma\rho
}{R^2}}). 
\label{th-sptorus} 
\end{equation} 
In both cases we define  
\begin{equation}
D_S(\sg) = -2 \frac{d \log K(\sg)}{d \log \sg},~~~{\rm i.e.}~~~
\lim_{\sg \to  0} D_S(\sg) = D_S. 
\end{equation} 
In  Fig.\ \ref{sp-theory-first} we show the behavior of $D_S(\sg)$ for
small $\sg$. In both cases $D_S(\sg)=2$ for small 
$\sg$ since the spacetime is two-dimensional. Further, in both
cases $D_S(\sg)$ eventually goes to zero simply because we have 
a finite spacetime volume and the large time diffusion will be 
dominated by the zero mode of the Laplacian, the constant mode, 
i.e.\ $K(\xi,\xi_0,\sg) \propto 1/V$ for  $\sg\to \infty$.
For a torus we see a non-trivial dependence on $\rho$.
This dependence is illustrated in Fig.\ \ref{sp-torus-fix}. 
For $R_2 >> R_1$ we see the appearance of a second
plateau at $D_S \approx 1$, reflecting the effective  
one-dimensional structure at a larger scale. 
Thus $D_S(\sg)$ contains information about the
spectral dimension ($D_S = 2$ in both cases due to the small 
$\sg$ behavior) and about the
large scale geometry of the manifold (obtained for intermediate $\sg$). 
Finally, the scaling for intermediate and large $\sg$ 
also provides information about $D_S$ since one in this region 
can obtain scaling by using the variable $\sg/V^{2/D_S}$ and  
comparing the diffusion for different spacetime volumes $V$. 

The matter action used in (\ref{action}) leads to a natural 
discretized version of the scalar Laplacian which appears in
eq.\ \rf{j2}. The spectral properties of
this operator can be analyzed in order to obtain additional 
information about the fractal properties of the ``blob''-geometry.
The diffusion process in the discrete diffusion-time 
variable $s$ defines  a probability field $P_i(s;T)$ for a particular
geometric configuration $T$ ($i$'s  label vertices in the triangulation $T$)
by the following equation, which is the discrete analogy of the 
standard diffusion equation \rf{j2}:
\begin{equation} 
P_i(s+1;T) = \frac{1}{3} \sum_{j\to i} P_j(s;T)
\label{diffusion} 
\end{equation} 
The summation is over the three  vertices $j$ neighboring vertex $i$. 
The initial condition for $s = 0$ is
$P_i(0;T) = \delta_{i,i_0}$, where typically $i_0$ is chosen 
to belong to spatial slice with the largest spatial volume. 
We measure the average return
probability $P(s) =\langle P_{i_0,i_0}(s;T)\rangle$ where the average is
taken over initial points $i_0$ and configurations $T$. The diffusion
process is performed on systems with a finite spacetime volume $N$ 
and again we study the limiting behavior and the scaling for large $N$. 
A function $D_S(s)$ which is the discretized analogy to \rf{dif1} is defined as
\begin{equation} 
D_S(s+1) = -2\log\left(\frac{P(s+2)}{P(s)}\right)/
\log\left(\frac{s+2}{s}\right).
\label{fractal} 
\end{equation}
We use the discretized logarithmic
derivative in the expression above in a form which permits to
discriminate between even and odd times $s$.

Below we show that we have qualitatively
similar behavior to $D_S(\sg)$ when we  measured
(\ref{fractal}) for CDT triangulations. 
Extracting information from such numerical ``measurements''
requires some care. We would like to 
define the spectral dimension $D_S$ as the limit of the 
function $D_S(s)$ for small $s$. However, for  very short time 
$s=1,2,\ldots,s_0$ we expect lattice artifacts to interfere
with the ``correct'' continuum result (below we will see 
that $s_0 \approx 200$).
For larger times the value of $D_S$  is seen as a plateau
of the function $D_S(s)$, but there will be  an uncertainty in 
identifying the plateau and extracting the corresponding value $D_S$. 
For even larger times we expect scaling similar
to that described for smooth 2d manifolds as a function of
the spacetime volume $N$, but for sufficient 
long diffusion times we enter a range where we may have a combination of
numerical uncertainty and finite-size effects related to the stalk/bulk
interference (in the case where we have a ``blob''). 

\begin{figure}[h] 
\begin{center}
\includegraphics[width=0.45\textwidth]{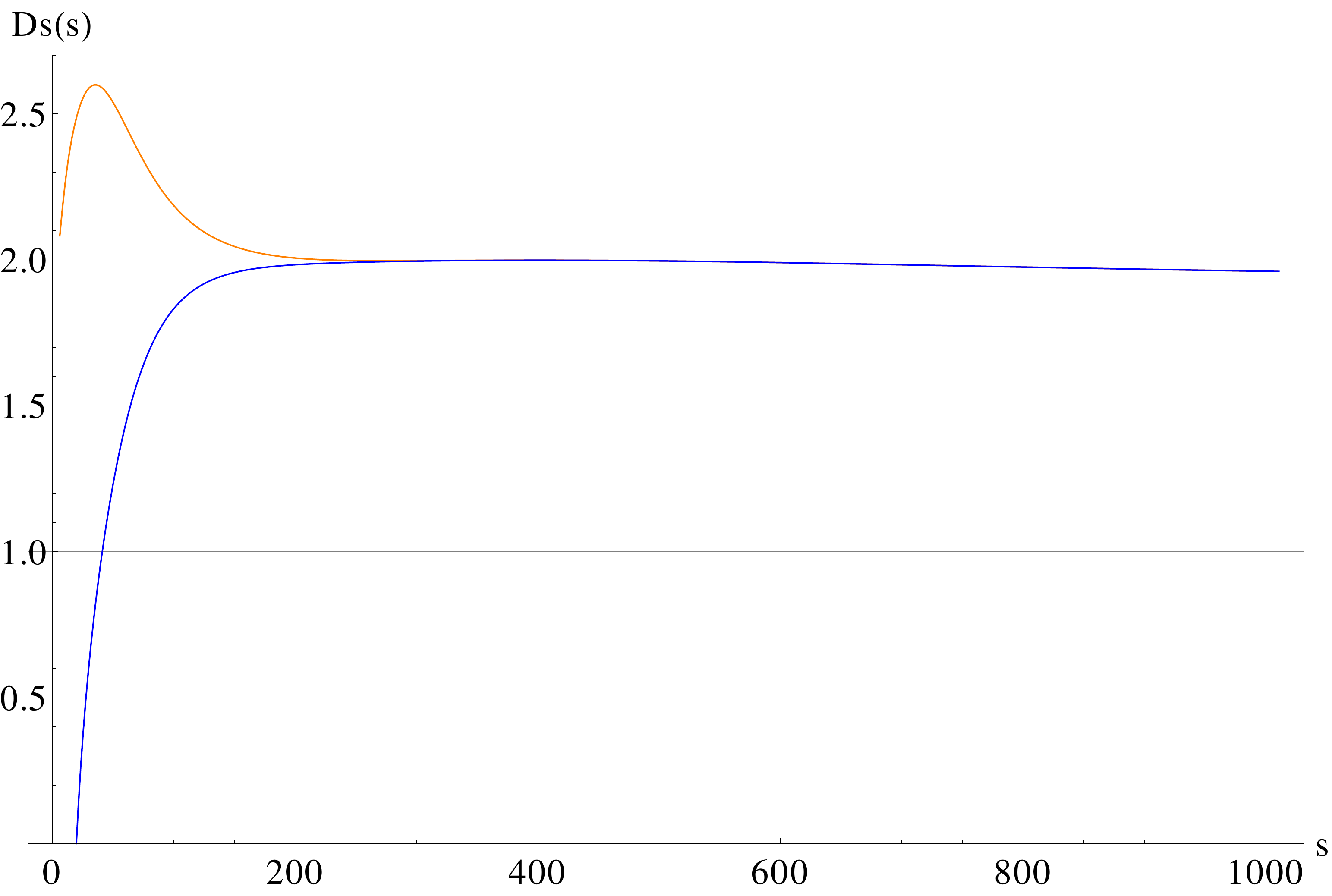}
\includegraphics[width=0.45\textwidth]{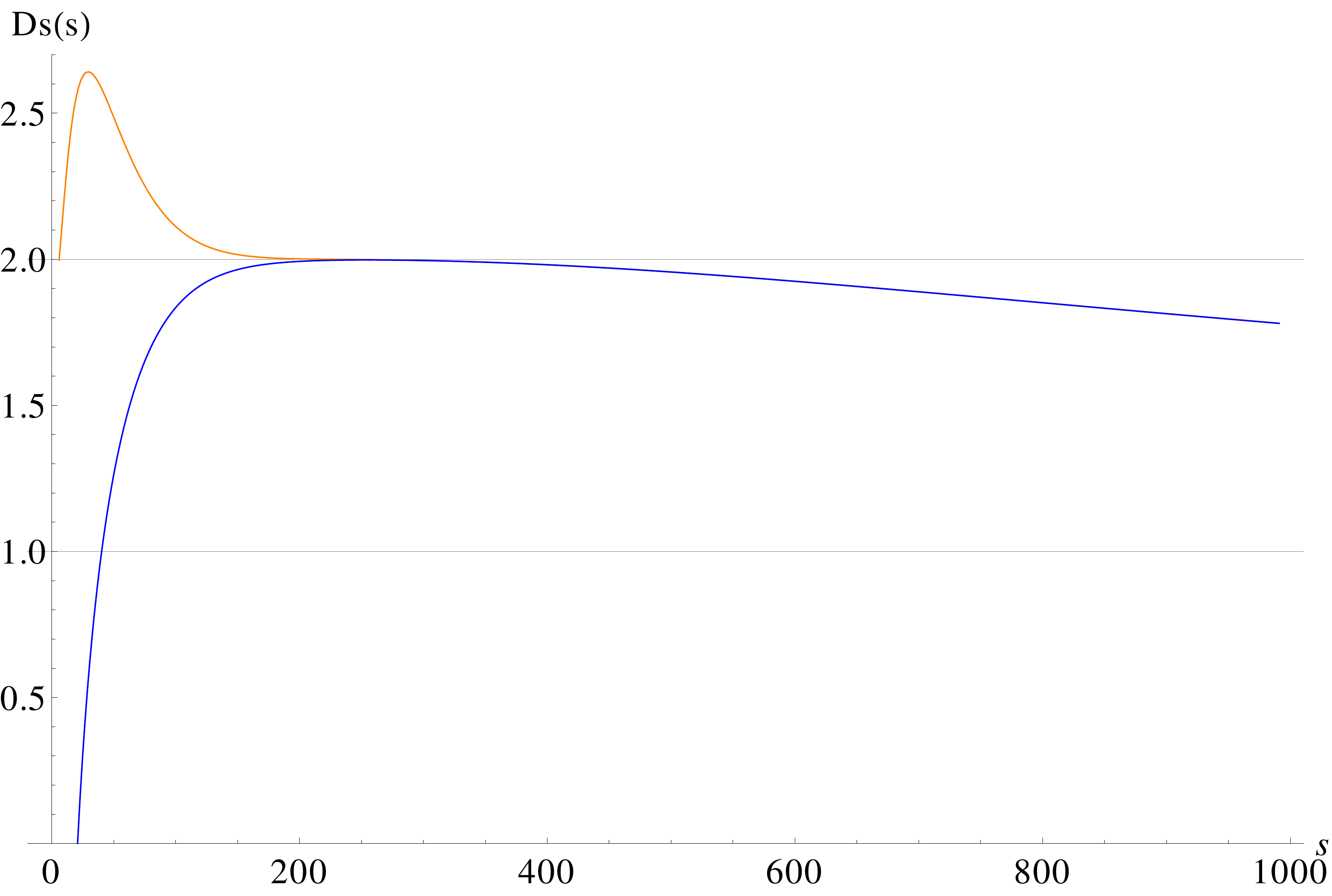} 
\end{center}
\caption{Dependence of $D_s(s)$ for $d=4$ (left) and pure gravity
(right) at small $s$,  $N=40000$, $L=200$ for both cases.}
\label{sp-simulation-first} 
\end{figure}

Let us first discuss the case $d=4$. In Fig.\
\ref{sp-simulation-first} we show the dependence of  $D_s(s)$
for small $s$ and we compare it with the pure gravity ($d=0$) behavior. 
In both cases the plot of $D_s(s)$ splits for very small $s$
into two curves, corresponding to  even and odd diffusion times. 
This is a typical lattice artifact. We can trust the estimate only for
times larger than the range of the even/odd split at $s\approx 200$.
For larger $s$ the plot  has a plateau corresponding to
$D_S \approx 2$, similar to that for a continuous two-dimensional geometry. 
We conclude that the value of the
spectral dimension is consistent with the canonical value $D_S = 2$ both
for $d=4$ and $d=0$. For $d=0$ this was expected 
since we have in  that case $D_H=2$ and since all indication
is that for $c_M \leq 1$ the CDT geometries can be considered as 
genuine two-dimensional without any significant fractal structure. 

\begin{figure}[h] 
\begin{center}
\includegraphics[width=0.45\textwidth]{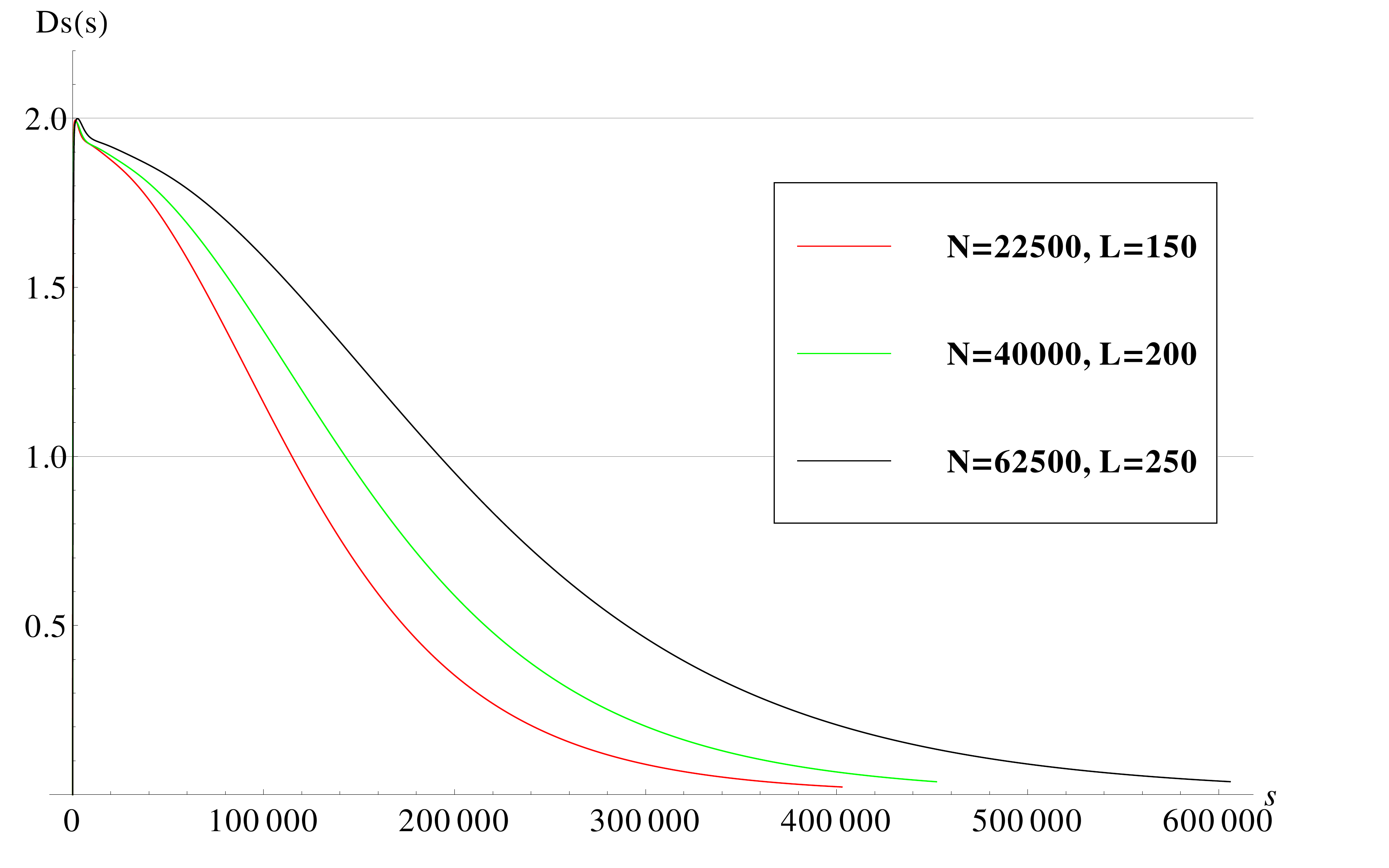}
\includegraphics[width=0.45\textwidth]{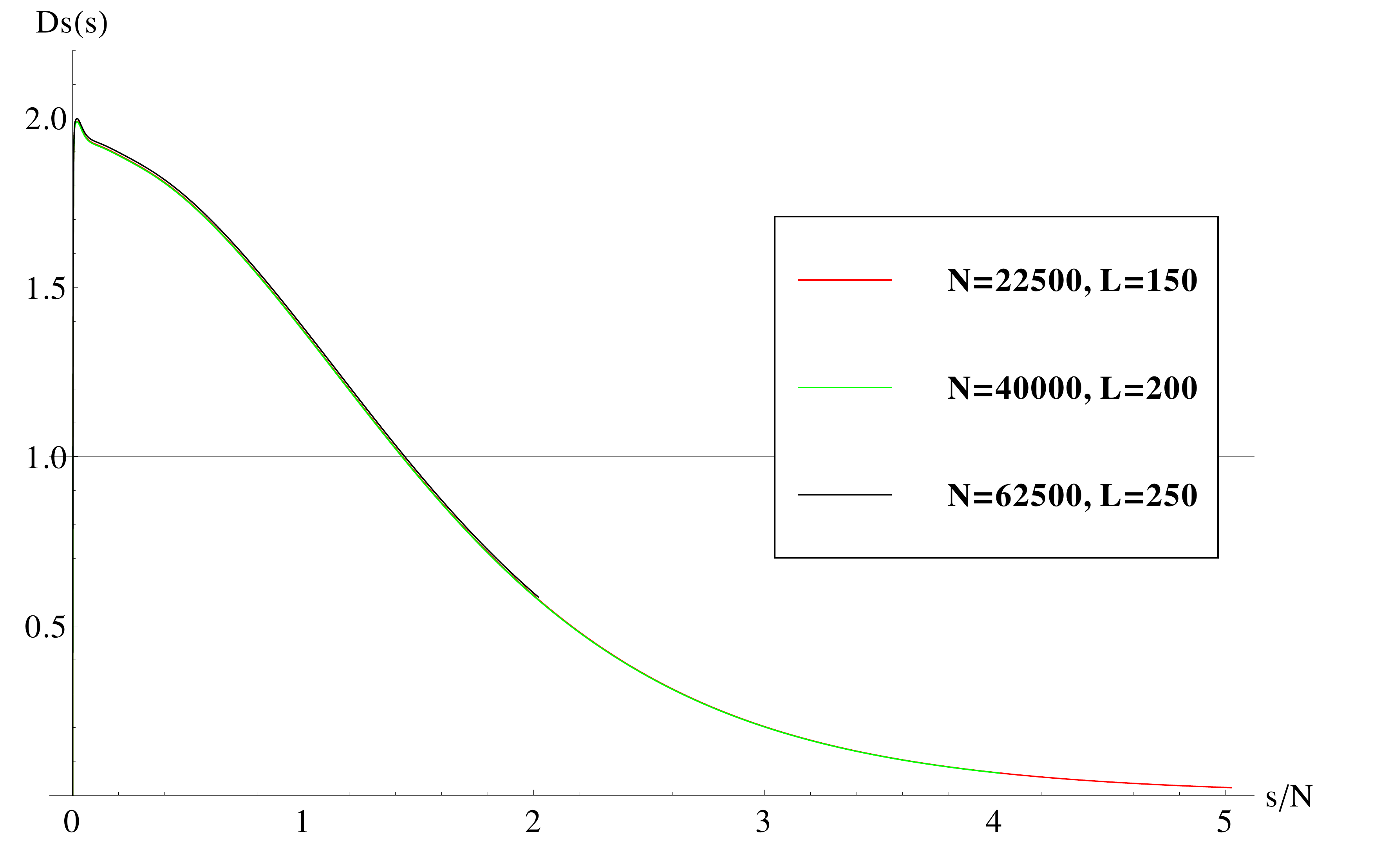}
\end{center} 
\caption{Scaling of $D_S(s)$ as a function of $s$ for $d=4$
and a number of volumes $N$. The left plot is not scaled, the right plot
is the same curve scaled as a function of $s/N$. }
\label{scaling-ds-massless} 
\end{figure}

\begin{figure}[h] \begin{center}
\includegraphics[width=0.45\textwidth]{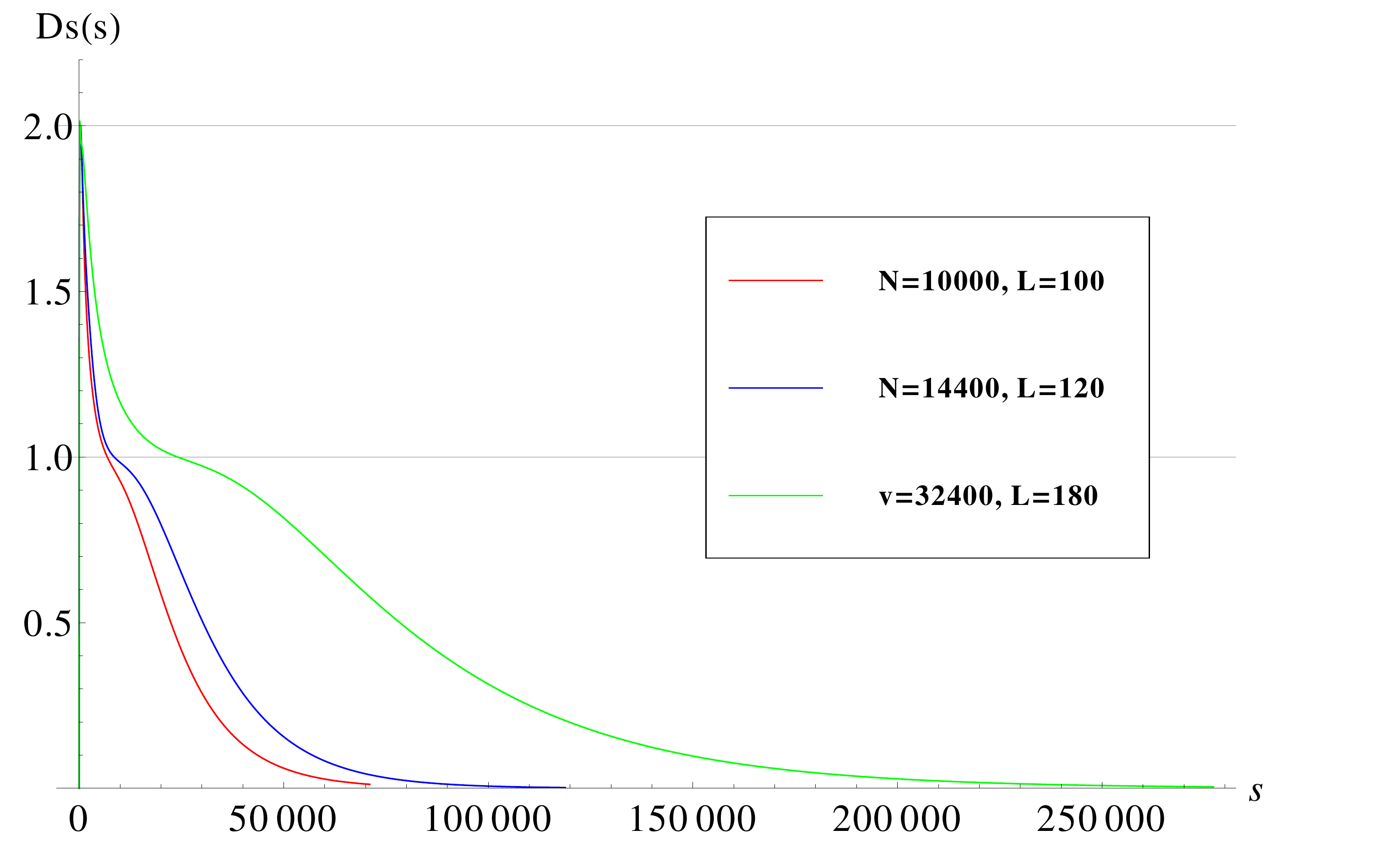}
\includegraphics[width=0.45\textwidth]{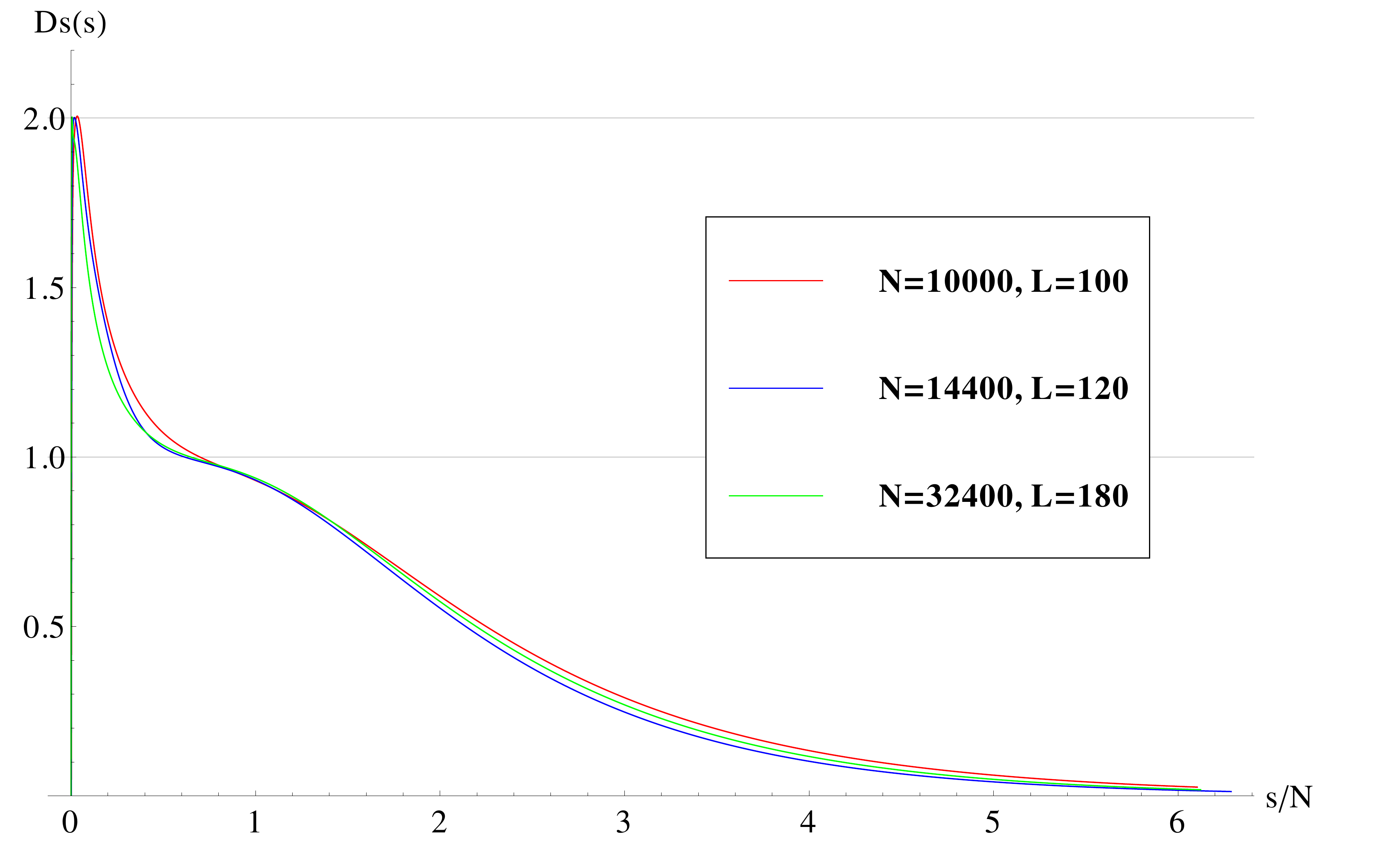} 
\end{center}
\caption{Scaling of $D_S(s)$ as a function of $s$ for pure gravity and a
number of volumes $N$.  The left plot is not scaled, the right plot is
the same curve scaled as a function of $s/N$. } 
\label{sp-pure}
\end{figure}

\begin{figure}[h] 
\begin{center}
\includegraphics[width=0.5\textwidth]{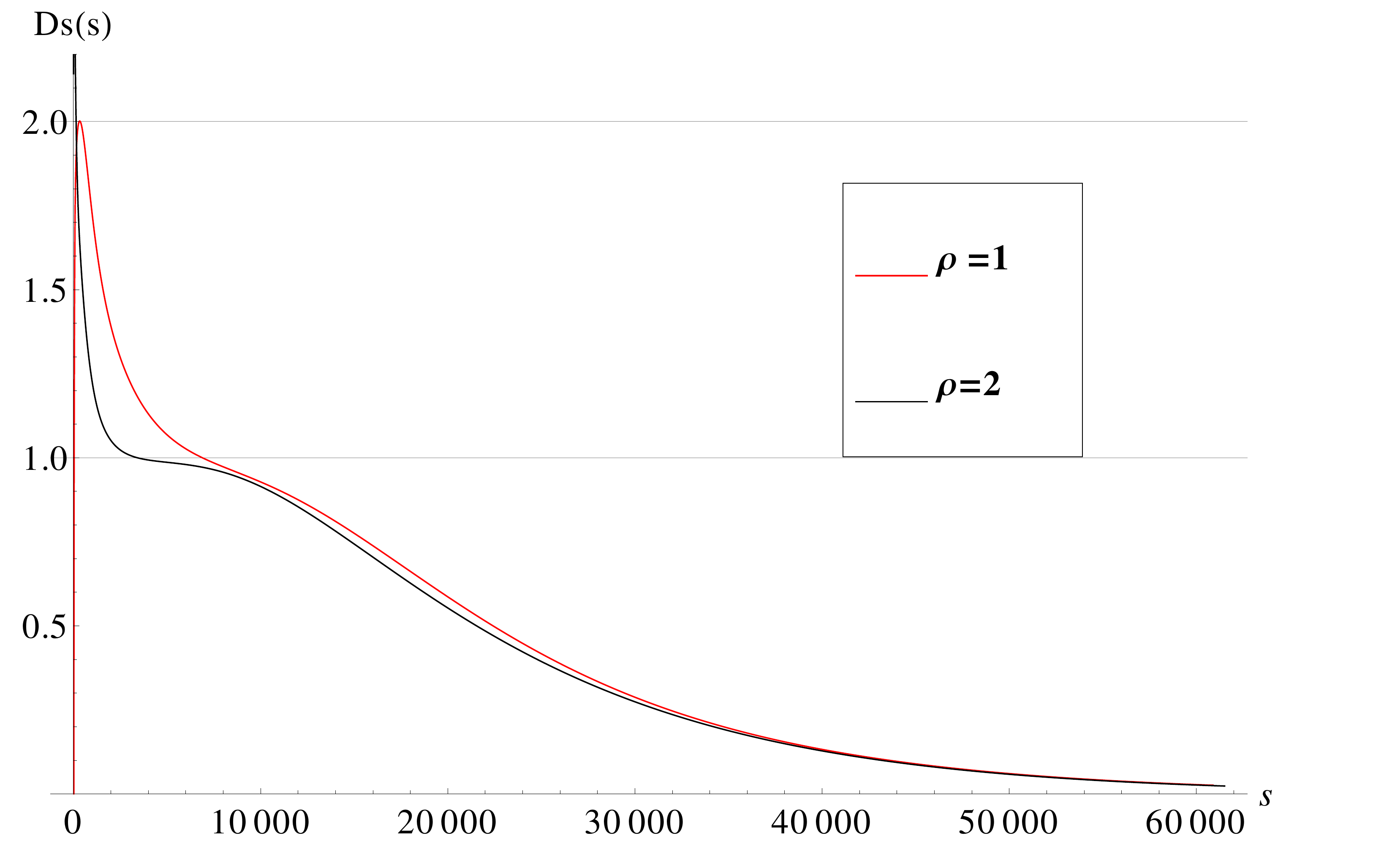} 
\end{center}
\caption{$D_S(s)$ for pure gravity: $\rho = 1$ and $\rho = 4$, we set $N = 10000$ for both cases and $L=100$ and $L=200$. } 
\label{pure-com}
\end{figure}

For $d=4$ it was not clear what to expect.
As mentioned above the Hausdorff dimension is in this case $D_H = 3$. 
The value $D_H=3$ is thus not transmitted to the spectral dimension $D_S$. 
The value $D_S = 2$ is further confirmed by looking at the scale 
dependence of $D_S(s)$.
In Fig.\  \ref{scaling-ds-massless} we plot $D_S(s)$
as a function of the  scaling variable $s/N^{2/D_S}$ with $D_S = 2$.
For the range of $s$ studied we see no trace of the toroidal geometry.
Diffusion is completely dominated by the ``blob''.
This is in  contrast to the  case of pure gravity ($d=0$). 
In Fig.\ \ref{sp-pure} we
show  $D_S(s)$ measured for $d=0$. It behaves qualitatively 
similar to  $D_S(s)$ calculated using the continuum diffusion 
equation  on a regular torus and it shows scaling when we use 
the re-scaled diffusion time $s/N$.
The plateau $D_S(s)=1$ as a function of $\rho$ is shown
in Fig.\  \ref{pure-com}. We see as expected that the 
plateau becomes more pronounced with increasing $\rho$. 


\section{Summary}\label{summary}

We have measured the spectral dimension of two-dimensional CDT 
quantum gravity coupled to four scalar fields. In this model  
spacetime scales anomalously, forming a blob where
the time scales as $N^{1/3}$ and space as $N^{2/3}$, $N$ being the 
volume of spacetime. We found that despite of this anomalous scaling,
indicating that the Hausdorff dimension $D_H=3$,  
the spectral dimension $D_S=2$, like for smooth two-dimensional
spacetimes\footnote{There exists other examples of a fractal 
geometry with $D_H \neq 2$ has $D_S=2$. The best know 
example is two-dimensional EDT coupled to conformal field theories. In this 
case a generic triangulation which appear in the path integral has a 
Hausdorff dimension $D_H(c) \neq 2$, depending on the central charge $c$
\cite{watabiki,kawai,aw,ab}. Nevertheless, the 
spectral dimension $D_S=2$ \cite{abw}.}. 
 
We compared this behavior with that of pure gravity (two-dimensional 
CDT without matter fields), 
where the Hausdorff dimension $D_H=2$ and where no blob is 
formed. Although this system
admits the existence of an analytic solution for some quantities of
interest, this is not the case for the heat kernel of the Laplacian,
which has to be studied numerically. We show that the spectral dimension
is $D_S = 2$, but in contrast to the $d=4$ case the diffusion 
would explore the full toroidal topology, and by varying the 
asymmetry parameter $\rho$ of the torus one could observe a secondary 
plateau at $D_S=1$, expressing the fact if one of the radii of the 
torus is small, long time diffusion will effectively see 
a one-dimensional thin tube.

\vspace{1cm}

\noindent {\bf Acknowledgments.} HZ is partly supported by the
International PhD Projects Programme of the Foundation for Polish
Science within the European Regional Development Fund of the European
Union, agreement no. MPD/2009/6.  JJ acknowledges the support of grant 
DEC-2012/06/A/ST2/00389
from the National Science Centre Poland.
JA and AG acknowledge support from the ERC-Advance grant 291092,
``Exploring the Quantum Universe'' (EQU). JA acknowledges support
of FNU, the Free Danish Research Council, from the grant
``quantum gravity and the role of black holes''.

\vspace{1cm}


 \end{document}